# Pliable Polaritons:

# Wannier Exciton-Plasmon Coupling in Metal-Semiconductor Structures

Jacob B Khurgin

Johns Hopkins University

Baltimore MD 21218 USA

**Abstract**

Plasmonic structures are known to support the modes with sub-wavelength volumes in which the field/matter interactions are greatly enhanced. Coupling between the molecular excitations and plasmons leading to formation of "plexcitons" has been investigated for a number of organic molecules. However, plasmon-exciton coupling in metal/semiconductor structures have not experienced the same degree of attention. In this work we show that the "very strong coupling" regime in which the Rabi energy exceeds the exciton binding energy is attainable in semiconductor-cladded plasmonic nanoparticles and leads to formation of Wannier Exciton-Plasmon Polariton (WEPP) that is bound to the metal nanoparticle and characterized by dramatically smaller (by factor of few) excitonic radius and correspondingly higher ionization energy. This higher ionization energy exceeding approaching 100meV for the CdS/Ag structures may make room temperature Bose Einstein condensation and polariton lasing in plasmonic/semiconductor structures possible.

## 1. Introduction

Polaritons [1] are bosonic quasi-particles formed by photons and matter excitations such as phonons [2], excitons [3, 4] , intersubband transitions [5], or others. Polaritons can be formed in a relative unconstrained geometry ("free space"), in the waveguides [6], or in the fully confining resonant structures as in the case of the widely studied exciton-photon polaritons in microcavities [4]. Polaritons are enabled by the coupling between the vacuum fluctuations of the electromagnetic field and the matter, quantified by the Rabi energy $\hbar\Omega$. When the Rabi energy exceeds the dissipation rates of both photons and excitons the exciton-photon system is said to enter the so-called "strong coupling regime"[7] and polaritons get formed as manifested by characteristic Rabi splitting. Furthermore, a number of fascinating phenomena takes place, most notably Bose Einstein condensation and polariton lasing. Satisfying the strong coupling criterion is far from being easy, and it is only in the last 20 years that the progress in micro cavity fabrication and growth of high quality III-V semiconductors has led to observation of large Rabi splitting, exciton condensation [8] and polariton lasing in multiple quantum wells (MQW) GaAs [9] or GaN [10] – based structures. A term "Ultrastrong coupling" (USC) had also been introduced to describe the situation where the coupling (Rabi) energy becomes comparable to the photon energy itself, meaning that the coupling changes electronic structure of the material itself. This regime is difficult to achieve in the optical range of frequencies but it can be feasible with THz radiation [11, 12]

At the same time, with the MQW exciton polariton in the semiconductor cavity another, intermediate, regime of the so-called "very strong coupling" (VSC) exits in which the Rabi energy $\hbar\Omega$ is much less than photon energy $\hbar\omega$ but it is comparable or larger than the exciton binding energy $\mathrm{E}_X$. Then as shown in [13]one can no longer consider coupling as a weak perturbation to the "rigid" exciton – the exciton in cavity polariton becomes flexible, or, better said, "pliable" as its radius decreases for the lower polariton and increases for the upper one causing increase of the effective binding energy in the lower polariton. The simple semi-analytical results of [13] had been confirmed by a rigorous Green function analysis in [14], and, after numerous attempts [15, 16], had been experimentally verified in [17].

Microcavities used in all the above experiments typically provide confinement in only one dimension which limits the strength of vacuum field that is inversely proportional to the square root of the cavity volume. Even three dimensional all dielectric cavities have volumes larger than $(\lambda/n)^3$ . Therefore, in recent years the attention of the polaritonic community had been turned in the direction of the plasmonic structures which enable localized surface plasmon (LSP) modes whose volume is far below the diffraction limit ($V<<(\lambda/n)^3$) [18] and in which electromagnetic field can be greatly enhanced in the vicinity of resonance. For this reason LSPs are widely used (or at least on their way to be used) in a variety of sensing [19], spectroscopic [20] and other applications [21]. Obviously vacuum field strength in LSP is also greatly enhanced which means that the coupling between LSP and the medium placed in close proximity to it also gets stronger, which results in strong Purcell enhancement of spontaneous radiation rate. Once the Rabi energy exceeds the damping (broadening) rate of LSP $\Gamma$, a new type of a quasiparticle –plasmon-exciton polariton or "plexciton" is being formed [22, 23]. Plexcitons had been observed with a variety of metals [24] [25]in the shape of both individual nanoparticles /nanoantennas and in their arrays [26]. Strong coupling has led to such exotic behavior as plexcitonic lasing [27] and condensation [28, 29]. There had been reports of USC regime [30] with Rabi splitting exceeding hundreds of meV [31, 32]

Most of the plexciton research had been performed with various types of organic and other molecules in which the exciton is essentially a local excitation localized on a scale of a few atoms and capable of moving around in the medium. One can easily model these excitons as isolated two-level entities that gets coupled via interaction with LSP cavity and thus forming plexcitons. The spatial characteristics of excitons do not change – one can say that these excitons, usually referred to as Frenkel excitons [33, 34] in condensed matter science, remain "rigid". But in the inorganic semiconductor medium excitons are formed from free carriers in the conduction and valence band attracted to each other, their wavefunctions are usually spaced over many lattice sites and they are usually referred to as Wannier excitons [35]. Radius of Wannier excitons can vary depending on the environment, and as have been shown in [13] it can change in the VSC regime. One can say that unlike rigid Frenkel excitons Wannier excitons are quite pliable. Furthermore, due to their lower mass Wannier excitons can travel through the medium easier than Frenkel excitons, and they can also be localized in the region where their potential energy is lower forming bound excitons.

In this paper we consider the interaction between Wannier excitons in the semiconductor and LSPs in metal nanoparticles forming a quasiparticle of a different flavor – Wannier Exciton Plasmon Polariton (WEPP). Using the example of WEPP in Ag/CdS structure. we show that VSC regime can be rather easily reached. In this VSC regime WEPP becomes quite flexible or pliable as the excitonic radius decreases at least threefold, and WEPP becomes bound to the Ag nanoparticle forming a shell with the thickness of a few nanometers.

## 2. Theory of WEPP

The Hamiltonian of the system consisting of a localized surface plasmon (LSP) and collective excitation of electron-hole pairs with momenta $\boldsymbol{k}_e$ and $\boldsymbol{k}_h$ forming Wannier exciton can be written as [36]

$$\hat{H} = \hbar\omega_{pl}\hat{a}^{\dagger}\hat{a} + \frac{1}{2}\sum_{\mathbf{k}_e,\mathbf{k}_h} \mathrm{E}_{\mathbf{k}_e,\mathbf{k}_h}\hat{b}^{\dagger}_{\mathbf{k}_e,\mathbf{k}_h}\hat{b}_{\mathbf{k}_e,\mathbf{k}_h} - \sum_{\mathbf{k}_e,\mathbf{k}_h} V_{c,\mathbf{k}_e\text{-}\mathbf{k}_h}\hat{b}_{\mathbf{k}_e,\mathbf{k}_h}\hat{b}^{\dagger}_{\mathbf{k}_e,\mathbf{k}_h} + iH_{\text{int}}\sum_{\mathbf{k}_e,\mathbf{k}_h}\left(\hat{a}\hat{b}^{\dagger}_{\mathbf{k}_e,\mathbf{k}_h} - \hat{a}^{\dagger}\hat{b}_{\mathbf{k}_e,\mathbf{k}_h}\right)$$

where $\hat{a}^{\dagger},\hat{a}$ and $\hat{b}^{\dagger}_{\mathbf{k}_e,\mathbf{k}_h},\hat{b}_{\mathbf{k}_e,\mathbf{k}_h}$ are the creation-annihilation operators for the plasmons with energy $\hbar\omega_{pl}$ and electron-hole pairs with energies $\mathrm{E}_{\mathbf{k}_e,\mathbf{k}_h}$, $V_{c,\mathbf{k}_e\text{-}\mathbf{k}_h} = e^2/\varepsilon_0\varepsilon_s\,|\boldsymbol{k}_e - \boldsymbol{k}_h|^2$ is Coulomb attraction energy, $\varepsilon_s$ is the static dielectric permittivity, and $H_{\text{int}}$ is the interaction energy between the plasmons electric field and the electron hole pair. The WEPP state can be described as

$$\left|P_{\alpha,\beta}\right\rangle = \alpha\left|1_{pl},0_{ex}\right\rangle + \beta\left|0_{pl},1_{ex}\right\rangle \tag{1}$$

where $\alpha$ and $\beta$ are the relative weight of plasmon and exciton (Hopfield coefficients [37]) , the ket vectors refer to the states with 1(0) plasmons and 0(1) excitons, the plasmon (or rather LSP) is described by its electric field $\boldsymbol{E}(\boldsymbol{R})$ and the Wannier exciton by its wavefunction $\Phi_{ex}(\boldsymbol{R},\boldsymbol{\rho})$ which can be approximated by the product of normalized wavefunctions for the center of mass (COM) motion $\Phi(\boldsymbol{R})$ and the relative motion of the electron and hole $\Psi(\boldsymbol{\rho})$. Here the relative electron-hole coordinate is $\boldsymbol{\rho} = \boldsymbol{r}_e - \boldsymbol{r}_h$ and the COM coordinate is $\boldsymbol{R} = \left(m_e\boldsymbol{r}_e + m_h\boldsymbol{r}_h\right)/M$ where $M = m_e + m_h$.

The total energy of the WEPP can then be written as

$$\mathrm{E}_{WEPP} = \left\langle P_{\alpha,\beta}\right|\hat{H}\left|P_{\alpha,\beta}\right\rangle = \alpha^2\hbar\omega_{pl} + \beta^2\mathrm{E}_{gap} + \beta^2\mathrm{E}_{cm} + \beta^2\mathrm{E}_{eh} - \beta^2\mathrm{E}_C - 2\alpha\beta\hbar\Omega \tag{2}$$

where $\mathrm{E}_{gap}$ is the bandgap energy, $\mathrm{E}_{cm} = -\hbar^2/2M\left\langle\Phi^*(\boldsymbol{R})\right|\nabla^2\left|\Phi(\boldsymbol{R})\right\rangle$ is the kinetic energy of the COM motion in exciton $\mathrm{E}_{eh} = -\hbar^2/2m_r\left\langle\Psi^*(\boldsymbol{\rho})\right|\nabla^2\left|\Psi(\boldsymbol{\rho})\right\rangle$ is the kinetic energy of the relative electron and hole motion, $m_r = (m_e^{-1} + m_h^{-1})^{-1}$ is reduced mass, $\mathrm{E}_C = -e^2\left\langle\Psi^*(\boldsymbol{\rho})\right|\rho^{-1}\left|\Psi(\boldsymbol{\rho})\right\rangle/4\pi\varepsilon_0\varepsilon_s$ is the Coulomb energy of electron-hole attraction, and $\hbar\Omega$ is the strength of exciton-plasmon coupling ($\Omega$ is a Rabi frequency). The strength of the coupling can be found as

$$\hbar\Omega = \left|\Psi(0)\int \Phi(\boldsymbol{R})\left[\boldsymbol{\mu}_{cv}\cdot\tfrac{1}{2}\boldsymbol{E}_{vac}(\boldsymbol{R})\right]d\boldsymbol{R}\right| \tag{3}$$

where the factor of ½ accounts for the fact only the positive frequencies component of the "vacuum" electric field $\boldsymbol{E}_{vac} \sim \cos\omega_{pl}t$ is responsible for the coupling between the states and $\boldsymbol{\mu}_{cv}$ is the matrix element of the dipole moment between the valence and conduction bands. It can be related to the interband matrix element of the momentum $P_{cv}$ as $\mu_{cv} = e\hbar P_{cv} / m_0 \mathrm{E}_{gap}$ where $\mathrm{E}_{gap}$ is the bandgap energy and then using the relation between $P_{cv}$ and the conduction band effective mass $m_0^2 / m_e = 2P_{cv}^2 / \mathrm{E}_{gap}$ [38] to obtain $\mu_{cv} = (e^2\hbar^2 / 2m_e\mathrm{E}_{gap})^{1/2}$.

Finding the "vacuum field" $\boldsymbol{E}_{vac}(\boldsymbol{R})$ of the SPP mode, responsible for vacuum Rabi oscillations and splitting is different from the case of dielectric micro cavities, where the total energy is equally split between electric and magnetic energies. In subwavelength LSP modes with characteristic size $s_{eff} << \lambda$ the magnetic energy $U_H$ is very small in comparison to the electric energy $U_E$ , roughly $U_H \sim (2\pi s_{eff} / \lambda)^2 U_E$ and can be neglected. Then the energy of the one LSP can be found as $\hbar\omega_{pl} = \frac{1}{4}\varepsilon_0 \int \left[\left(\partial(\omega\varepsilon_r)/\partial\omega\right)| E_{vac}(\boldsymbol{R})|^2\right] d\boldsymbol{R}$ where $\varepsilon_r$ is the co-ordinate dependent relative permittivity at optical frequencies. Furthermore, in the metal described by the Drude relative permittivity $\varepsilon_m(\omega) = 1 - \omega_p^2/\omega^2$ , $\partial(\omega\varepsilon_m)/\partial\omega \equiv \omega_p^2/\omega^2$ and as shown in [39, 40] in the absence of the magnetic field exactly one half of the total energy of the plasmon is contained in the form of kinetic energy of the free electrons, $U_K = \frac{1}{2}U_E = \frac{1}{4}\varepsilon_0 \int_{metal} (\omega_p^2/\omega^2)\left[| E_{vac}(\boldsymbol{R})|^2\right] d\boldsymbol{R}$ . It follows than that $\hbar\omega_{pl} = \frac{1}{2}\varepsilon_0 \int_{metal}\left[(1-\varepsilon_m)| E_{vac}(\boldsymbol{R})|^2\right] d\boldsymbol{R}$. Let us normalize the field of the plasmonic mode to the field at some specific point, for example point $\boldsymbol{R}_0$ where the electric field is maximum, $\hat{\boldsymbol{e}}(\boldsymbol{R}) = \boldsymbol{E}(\boldsymbol{R}) / E(R_0)$ and then we immediately obtain

$$\tfrac{1}{2}\boldsymbol{E}_{vac}(\boldsymbol{R}) = \left(\frac{\hbar\omega_{pl}}{2\varepsilon_0 V_{pl}}\right)^{1/2} \hat{\boldsymbol{e}}(\boldsymbol{R}) \tag{4}$$

where the "effective volume" of the plasmon is

$$V_{pl} = \int_{metal}\left[(1-\varepsilon_m)|\hat{e}(\boldsymbol{R})|^2\right] d\boldsymbol{R} \tag{5}$$

Note that the coefficient in front of $\hat{\boldsymbol{e}}(\boldsymbol{R})$ in (4) is precisely the coefficient relating operator of the electric field and the creation/annihilation operators [41] for the modes in vacuum with the only difference being the definition of the mode volume (5).

Substitute (5) into (3) and obtain

$$\hbar\Omega = \left(\frac{\hbar\omega_{pl}}{2\varepsilon_0 V_{pl}}\right)^{1/2} \Psi(0) \left(\frac{e^2\hbar^2}{2m_e\mathrm{E}_{gap}}\right)^{1/2} \int \Phi(\boldsymbol{R})\left[\hat{\boldsymbol{\mu}}_{cv}\cdot\hat{\boldsymbol{e}}(\boldsymbol{R})\right]d\boldsymbol{R} \tag{6}$$

where $\hat{\boldsymbol{\mu}}_{cv}$ is the unit vector collinear with the momentum of the interband transition. Integration over the volume involves averaging over the direction so the mean value of the projection of $\hat{\boldsymbol{\mu}}_{cv}$ onto the direction of electric field is $1/\sqrt{3}$ . Then we introduce the effective overlap between the plasmon electric field and the exciton wavefunction as $\kappa = V_{pl}^{1/2}\int \Phi(\boldsymbol{R})\hat{e}(\boldsymbol{R})d\boldsymbol{R}$ and finally obtain (assuming $\hbar\omega_{pl} \sim \mathrm{E}_{gap}$)

$$\hbar\Omega = \frac{1}{2\sqrt{3}}\left(\frac{e^2\hbar^2}{m_e\varepsilon_0}\right)^{1/2}\kappa\Psi(0) \tag{7}$$

Finally we can normalize the dimensions to the free exciton radius $a_0 = 4\pi\varepsilon_0\varepsilon_s\hbar^2/m_r e^2$ and the energy to the binding (Rydberg) energy of 1S exciton $\mathrm{E}_R = m_r e^4/32\pi^2\hbar^2\varepsilon_0^2\varepsilon_s^2$ where $\varepsilon_s$ is the static dielectric constant of the semiconductor and obtain

$$\hbar\Omega = \frac{2\pi^{1/2}}{\sqrt{3}}\mathrm{E}_R a_0^{3/2}\left(\varepsilon_s\frac{m_r}{m_e}\right)^{1/2}\kappa\Psi(0), \tag{8}$$

and also

$$\begin{aligned}
\mathrm{E}_{eh} &= -\mathrm{E}_R a_0^2\left\langle \Psi^*(\boldsymbol{\rho})\middle|\nabla^2\middle|\Psi(\boldsymbol{\rho})\right\rangle \\
\mathrm{E}_{cm} &= -\mathrm{E}_R a_0^2\left\langle \Phi^*(\boldsymbol{R})\middle|\nabla^2\middle|\Phi(\boldsymbol{R})\right\rangle m_r/M \\
\mathrm{E}_C &= -2\mathrm{E}_R a_0\left\langle \Psi^*(\boldsymbol{\rho})\middle|\rho^{-1}\middle|\Psi(\boldsymbol{\rho})\right\rangle
\end{aligned} \tag{9}$$

The wavefunction for the relative electron-hole motion is that of the usual lowest Hydrogen state $\Psi(\rho) = \exp(-\rho/a)/\pi^{1/2}a^{3/2}$ Therefore, if all the dimensions are normalized to the Bohr radius of exciton and the energy to that of binding energy we obtain

$$\mathrm{E}_{WEPP} = \alpha^2\Delta_{pl} - \beta^2\left\langle \Phi^*(\boldsymbol{R})\middle|\nabla^2\middle|\Phi(\boldsymbol{R})\right\rangle m_r/M + \frac{\beta^2}{a^2} - 2\frac{\beta^2}{a} - \frac{4}{\sqrt{3}}\alpha\beta\left(\varepsilon_s\frac{m_r}{m_e}\right)^{1/2}\frac{\kappa}{a^{3/2}} \tag{10}$$

where $\Delta_{pl} = \left(\hbar\omega_{pl} - \mathrm{E}_{gap}\right)/\mathrm{E}_R$ and the energy $\mathrm{E}_{WEPP}$ is now measured relative to the bandgap.

## 3. WEPP bound to a metal nanosphere

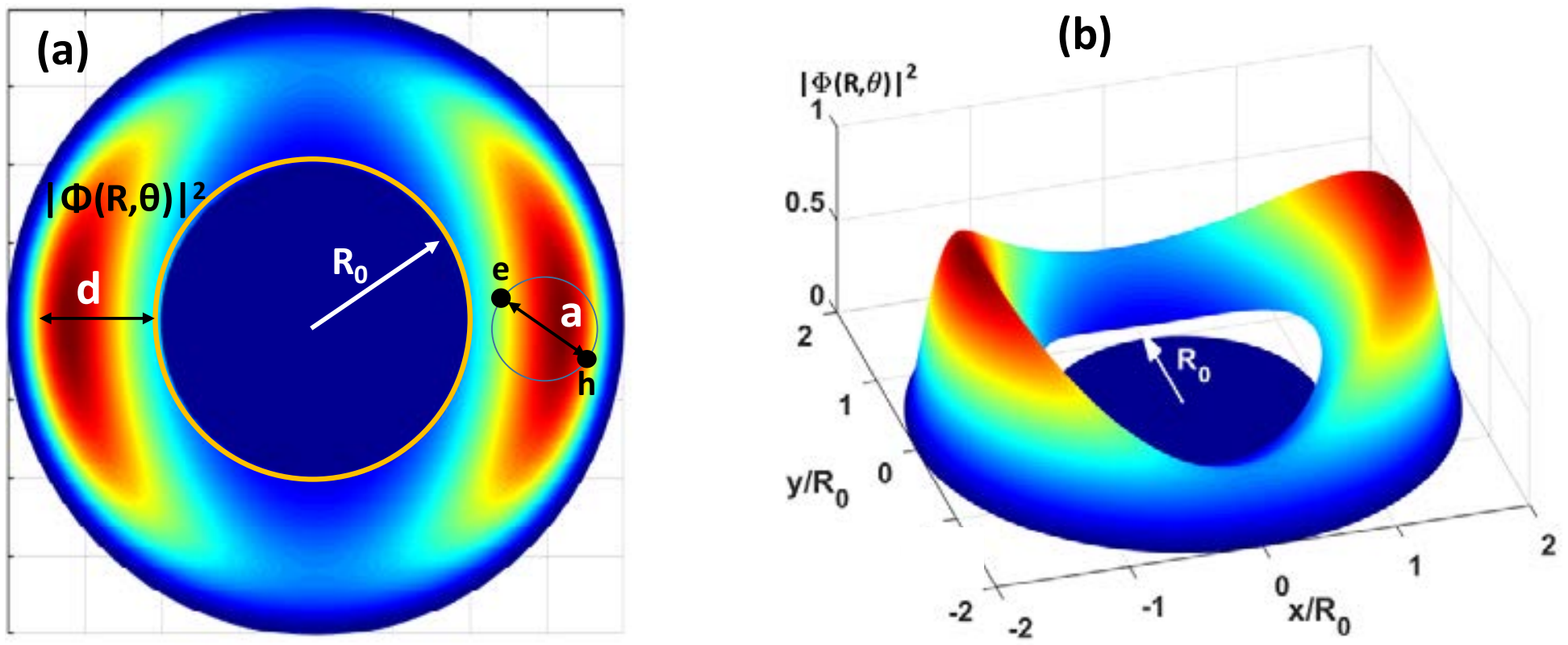

**Figure 1** (a) Wannier exciton localization of the on the metal nanosphere of radius $R_0$. $|\Phi(R,\theta)|^2$ is the probability density for the exciton center of mass and *a* is the excitonic radius. (b) 3-dimensional representation of center of mass probability density around metal nanospheres

To continue analysis, it is necessary at this point to choose a particular geometry. In order to facilitate carrying on analytical derivations for as long as possible before finally switching to numerical calculations, we consider the most simple example of a spherical metal nanoparticle with radius $R_0$ a s shown in Fig.1a. These nanospheres are capable of supporting a strong dipole surface plasmon mode whose electric field magnitude relative to the maximum field just outside the nanoparticle ( $R = R_0$ ) is

$$\hat{e}(\boldsymbol{R}) = \begin{cases} \frac{1}{2}\sqrt{3\cos^2\theta + 1}(R_0 / R)^3 & R > R_0 \\ \frac{1}{2} & R < R_0 \end{cases} \tag{11}$$

The condition for the surface plasmon resonance is $\varepsilon_m(\omega_{pl}) = -2\varepsilon_{s,opt}$ where $\varepsilon_{s,opt}$ is the relative dielectric permittivity of semiconductor cladding at optical frequencies whose dispersion (not counting the exciton!) is insignificant over the range of frequencies considered here. Therefore from (5) $V_{pl} = \frac{1}{3}\pi R_0^3(1+2\varepsilon_{s,opt})$. To avoid the surface recombination of excitons the wide gap spacer of a few Angstrom thickness must surround the metal as shown in Fig.1.a That will cause some distortion of the electric field due to the difference between the permittivities, but this difference can be ignored in the present order-of magnitude estimate.

Next we consider the COM wavefunction of the bound exciton $\Phi(\boldsymbol{R}) = \Phi_R(R)\Phi_\theta(\theta)$. To start with we assume that the angular function of the exciton is the same as the field, $\Phi_\theta(\theta) = \frac{1}{2}\sqrt{3\cos^2\theta + 1} / \sqrt{2\pi}$, where $(2\pi)^{-1/2}$ stand for normalization in the angular domain. Then the angular integral in the expression for $\kappa$ is $2\pi\int_{-1}^{1}\hat{e}(\theta)\Phi_\theta(\theta)\sin(\theta)d\theta = \sqrt{2\pi}$ and $\kappa = \sqrt{6/(1+2\varepsilon_{s,opt})}F_\kappa$, where $F_\kappa = R_0^{3/2}\int \Phi_R(R)R^{-1}dR$.We now find the kinetic energy of the angular motion

$$\mathrm{E}_{kin,\theta}(R) = \left\langle \Phi_\theta(\theta) \right| \nabla_\theta^2 \left| \Phi_\theta(\theta) \right\rangle = \left\langle \Phi_\theta(\theta) \right| \frac{1}{R^2}\frac{1}{\sin\theta}\frac{\partial}{\partial\theta}\frac{1}{\sin\theta}\frac{\partial}{\partial\theta}\left| \Phi_\theta(\theta) \right\rangle = \frac{\pi}{2\sqrt{3}R^2} \qquad (12)$$

Next the radial wavefunction must be chosen, and it must satisfy the boundary condition $\Phi_R(R_0) = 0$ hence an appropriate radiation wavefunction can be

$$\Phi_R(R) = \frac{2}{d^{3/2}(R_0^2 + 3R_0 d + 3d^2)^{1/2}}(R - R_0)e^{-(R-R_0)/d}, \qquad (13)$$

where $d$ is the distance of the peak from the nanosphere surface, to which we can refer to as "polariton shell thickness.

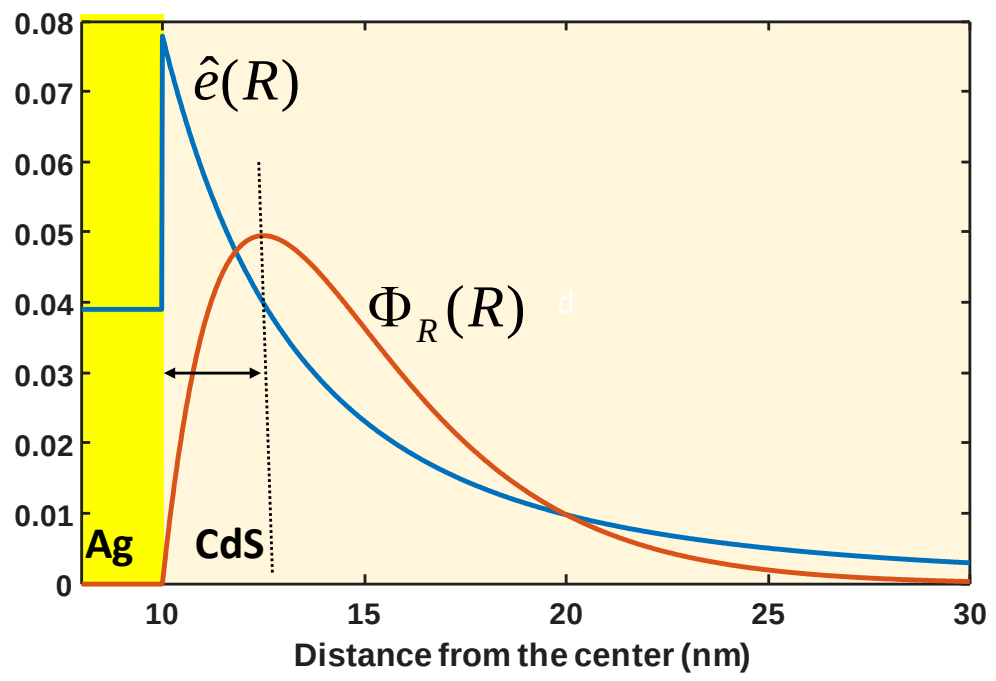

**Figure 2** Overlap between the electric field distribution of the plasmon mode $\hat{e}(R)$ and center-of-mass exciton wavefunction $\Phi_R(R)$ of the Wannier Exciton Plasmon Polariton (WEPP).

This wavefunction is plotted in the curve Fig. 2 next to the normalized electric field distribution, while the total COM probability density $\left|\Phi_R(R)\Phi_\theta(\theta)\right|^2$ is shown in Fig.1 a and b. Now we calculate the all-important overlap integral in

$$F_\kappa(d/R_0) = \frac{2}{[(d/R_0) + 3(d/R_0)^2 + 3(d/R_0)^3]^{1/2}}\left[1 - \exp^{R_0/d}\frac{R_0}{d}Ei_1(R_0/d)\right] \qquad (14)$$

where $Ei_1(x) = \int_1^\infty \exp(-tz)/t dt$ is the exponential integral. The function is shown in Fig.3a and it reaches the maximum value of roughly 0.5 for $d \sim 0.22R_0$

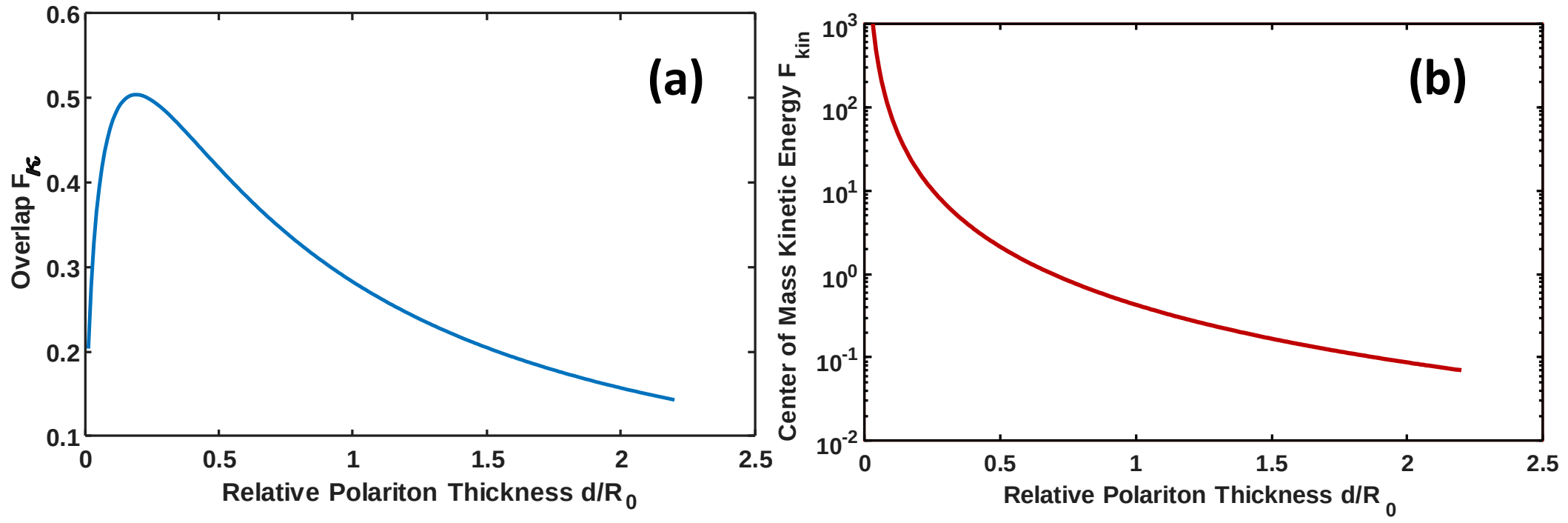

Figure 3. Dependences of (a) the overlap between the exciton and plasmon $F_\kappa$ and (b) center of mass kinetic energy $F_{kin}$ on relative thickness of the polariton shell $d/R_0$

Now all that is left is to calculate the kinetic energy term $\left\langle \Phi_R^*(\boldsymbol{R}) \right| -\nabla^2 \left| \Phi_R(\boldsymbol{R}) \right\rangle = F_{kin}(d/R_0)/R_0^2$, where

$$F_{kin}(d/R_0) = \frac{(R_0/d)^2 + 2(R_0/d) + \pi/2\sqrt{3}}{3(d/R_0)^2 + 3(d/R_0) + 1} \tag{15}$$

is shown in Fig. 3b. Now, the equation for the total lower WEPP energy follows from (10)

$$\mathrm{E}_P = \alpha^2 \Delta_{pl} + \frac{\beta^2}{R_0^2}\frac{m_r}{M} F_{kin}(d/R_0) + \frac{\beta^2}{a^2} - 2\frac{\beta^2}{a} - \frac{4\alpha\beta}{a^{3/2}}\sqrt{\frac{m_r}{m_e}\frac{2\varepsilon_s}{1+2\varepsilon_{s,opt}}} F_\kappa(d/R_0) \tag{16}$$

Clearly, for large nanosphere radius $R_0$ the exciton polariton will probably localize within the thin shell $d \sim 0.22R_0$, but for smaller radii the COM kinetic energy term, inversely proportional to $R_0^2$ may become too large and it will force the polariton shell thickness to increase and thus reduce the coupling strength $\hbar\Omega$. As far as the excitonic radius, it is determined by the interplay of the last three terms in (16) where positive kinetic energy term $\sim a^{-2}$ is counteracted by the Coulomb term $\sim a^{-1}$ and Rabi splitting $\sim a^{-3/2}$.

## 4. Material Choice and Background Oscillation Strength

We shall now consider a choice of materials. The metal sphere can be made of silver and among the semiconductors cadmium sulfide (*CdS*) appears to be a good candidate because it supports robust free Wannier excitons [42, 43] with radius of $a_0 \sim 3nm$ and binding energy of $\mathrm{E}_R = 30meV$ observable at room temperature. Furthermore, a surface plasmon supported by a small *Ag* nanosphere surrounded by *CdS* has energy of $\hbar\omega_{pl} \sim 2.30eV$, which places it in the region where intrinsic losses in silver are small and also very close to the bandgap exciton energy 2.42 *eV*. It means that detuning $\Delta_{pl}$ can be always adjusted within a fairly broad range of positive and negative values by either small variations in the *Ag* nanoparticle shape (making it slightly oblong) or changing semiconductor composition by introducing small fraction of selenium[44] to shift the bandgap energy of $CdS_{1-x}Se_x$ slightly downward.

Interestingly, given the relevant material characteristics of *CdS*: $m_e = 0.2m_0$ , $m_v = 0.8m_0$, $\varepsilon_s = 8.9$ and $\varepsilon_{s,opt}^0 = 6.35$, the square root term in (16) is equal to 1.02 , i.e. very close to unity . In fact, this term is not that far from unity for any polar semiconductor in which $\varepsilon_s > \varepsilon_{s,opt}$ and $m_e << m_v$ (It is 1.1 for *GaN*, for instance and 1.01 for *GaAs*) .

If we compare (16) with the corresponding equation for VSC in the microresonators incorporating QW's [13], two major differences can be identified. First one is that the non-uniform electric field in the plasmon mode causes exciton-polariton to become bound and introduces the third (in addition to Hopfield coefficients and exciton radius) variational parameter –shell thickness *d*. Second difference is the three-dimensional character of exciton which is manifested in the presence of $a^{-3/2}$ in place of $a^{-1}$ in the last term in (16) . This stronger dependence of oscillator strength and Rabi energy tends to reduce the excitonic radius in CdS WEPP from already small 2.8 nm by at least an order of magnitude, where it practically becomes a Frenkel exciton with a very large oscillator strength [45]. But this giant oscillator strength of the exciton with a small radius does not appear from nowhere. Since the exciton wavefunction is simply a superposition of the free carrier states in valence and conduction bands, as the oscillator strength of exciton grows with reduction of its radius, the sum of the oscillator strengths of continuum interband transitions decreases. And it is these transitions that are primarily responsible for the susceptibility $\varepsilon_{opt} - 1$ at optical frequencies. Therefore, when the radius of the exciton approaches the radius of the unit cell, $r_0$, essentially the entire oscillator strength of the interband transitions gets transferred to the exciton and the susceptibility decreases. The reduction of "background" optical dielectric constant can be approximated as $\varepsilon_{s,opt}(a) = \varepsilon_{s,opt}^0 - \left(\varepsilon_{s,opt}^0 - 1\right) r_0^3 / a^3$ . This assures that the same electron-hole pair transitions are not counted twice – first as the free carrier transition and then again as constituents of excitonic transition. This reduction will case a small change in the coupling strength term under the square root in (16) but the main effect of it will be in upward shift of the "undressed", i.e. in absence of exciton frequency of the plasmonic mode which can be found as

$$\Delta_{pl}(a) = \Delta_{pl}^0 + \hbar\omega_p \left[ \left(1 + 2\varepsilon_{s,opt}(a)\right)^{-1/2} - \left(1 + 2\varepsilon_{s,opt}^0\right)^{-1/2} \right] \quad (17)$$

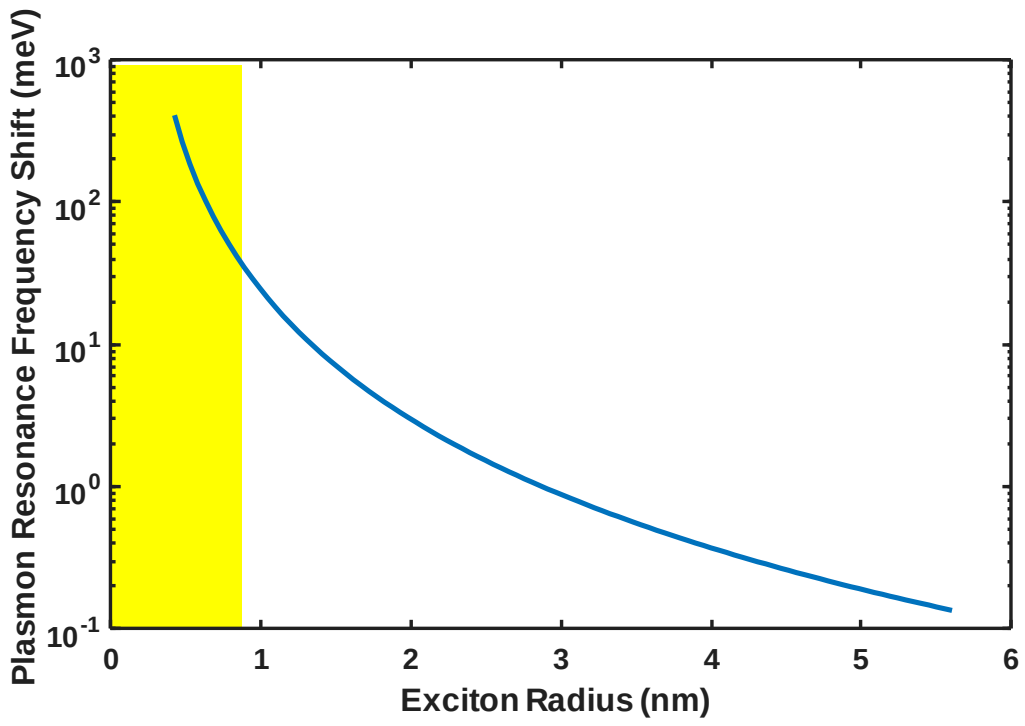

**Figure 4** Upward shift of the "undressed LSP resonance" as a function of the exciton radius *a*. Shaded region indicate the range of excitonic radii at which the upward shift of LSP precludes its effective coupling to the exciton.

As shown in Fig. 4 the reduction of exciton radius means that the oscillator strength of the free carrier interband transitions gets reduced which causes the upward shift of the "undressed" plasmon which prevents coupling with the exciton into polariton mode. While the exact form of $\varepsilon_{opt}(a)$ dependence is difficult to obtain, our approximation works really well in assuring that the exciton radius can only approach the unit cell size but never get smaller than that as indicated by the shaded region in Fig.4.

Introduction of dependence of plasmon resonance on excitonic radius means that not only VSC between the exciton and plasmon affects the exciton energy and shape but it also affects (albeit less strongly) the energy and shape of plasmon. This is to be expected since (as will be discussed below) formation of polariton is essentially a classical phenomenon. From this point of view, the "mobile" polarizable medium is attracted towards the electromagnetic mode (which is manifested by the phenomena taking place in optical tweezers [46] and multitude of micro opto-mechanical devices [47] ). Therefore, the easily polarizable and very mobile exciton is naturally attracted towards the plasmon mode. At the same, if the polarizable medium is not mobile, then it is the electromagnetic mode itself that tends to gravitate towards the more polarizable medium – hence all the waveguiding and other light-confining phenomena in dielectrics. It is only natural then to expect that the shape and energy of the plasmon mode will also change in the presence of polarizable medium. Interestingly, in most situations one side of the light-polarizable medium interaction easily dominates the other – in optical tweezers, for instance the micro-particles move while the optical mode stays largely undistorted. In the optical waveguides the situation is exactly the opposite – the light "moves" while the medium stays put. What makes the case of WEPP special is the fact that both optical mode and the exciton move and neither motion should be neglected.

## 5. Results

We can now perform minimization of plasmon-exciton-polariton energy (16) over three variational parameters, $\alpha$, $d$, and $a$ for different nanoparticle sizes $R_0$ and detuning between the bandgap (in the absence of exciton) and plasmon mode $\Delta_{pl}^0$. Note that unlike typical dispersion curves for cavity polaritons [3, 4] where change in detuning is achieved simply by changing the in-plane kinetic energy of COM motion (i.e. the angle of incidence) in real time, for bound exciton-polaritons each data point corresponds to the change of either semiconductor composition or the aspect ratio of Ag nanoparticle. The entire span of detunings of $\pm 200meV$ can be handled by growing $Zn_xCd_{1-x}S$ [48] for positive detuning and $CdS_{1-x}Se_x$ [44] for negative ones with the alloy fraction $x$ not exceeding 30%. . Of course for the experimental purposes one should consider only a few compositions where the Rabi splitting is maximum and dressed exciton radius is small, as shown in the results are displayed in Figs 5-7.

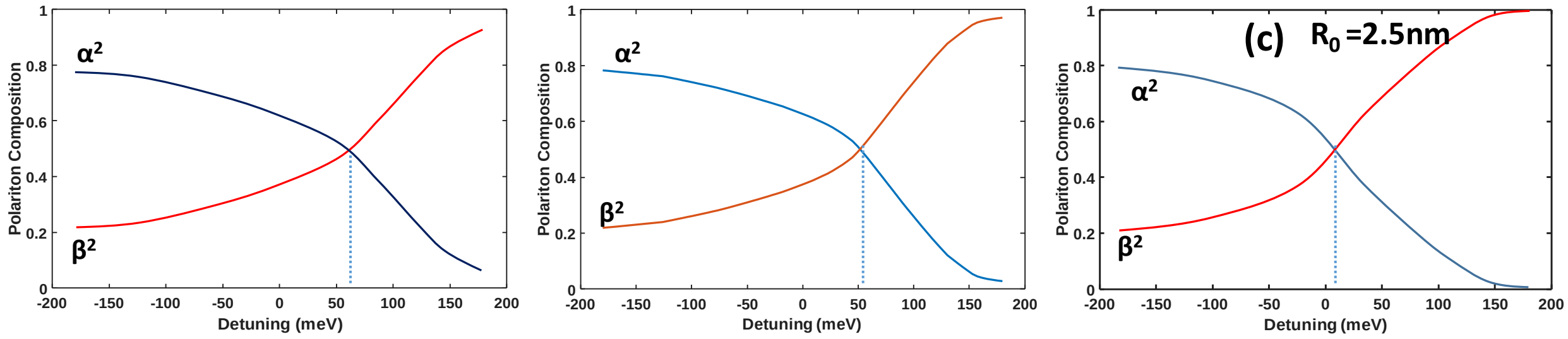

**Figure 5.** Polariton composition ($\alpha^2$ is plasmonic and $\beta^2$ is excitonic weight) versus the energy of the "undressed" surface plasmon relative to the bandgap energy for three different radii of the Ag nanosphere.

First, in Fig.5 the plasmon and exciton weights $|\alpha|^2$ and $|\beta|^2$ in the WEPP state are shown for three different values of the nanosphere radius $R_0$. One can see a substantial difference between these curves and what one would expect from the case where the plasmon mode interacts with the "rigid exciton" that can be described by the two level system. The curves in Fig.5 are not symmetric around the resonance, especially for the larger nanospheres of Fig.5a and b. While plasmon resonance stays well below the bandgap the WEPP composition varies slow as its character gradually changes from 100% plasmonic to being a partially excitonic. But once the 50%/50% composition is achieved the character of lower WEPP quickly becomes mostly excitonic. Note that at positive detunings "rigid" polariton is expected to have mostly excitonic character $\alpha < \beta$ however in VSC regime adding more of plasmonic (increased $\alpha$ ) character causes the energy lowering and as a result 50%/50% split occurs at positive detunings.

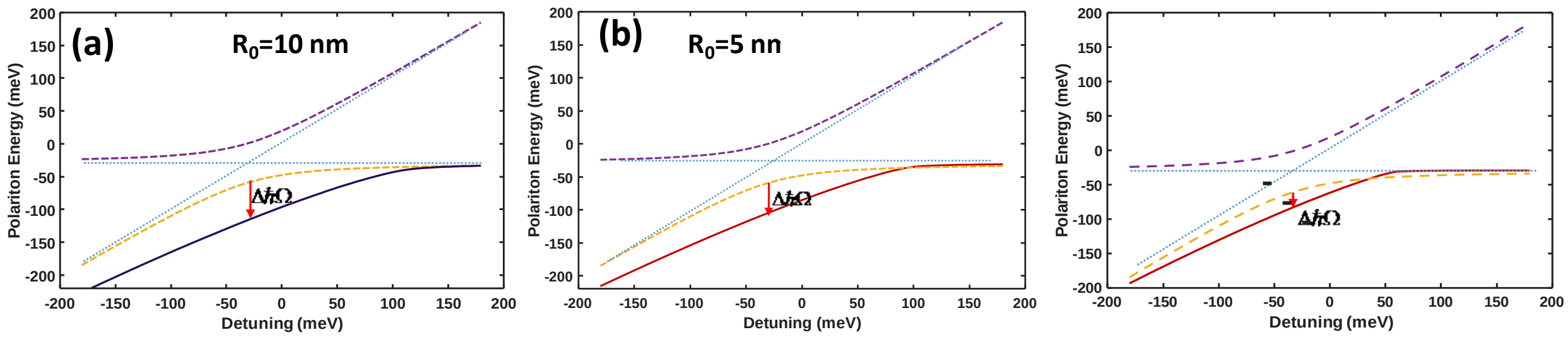

**Figure 6.** Lower polariton dispersion versus the energy of the "undressed" surface plasmon relative to the bandgap energy for three different radii of the Ag nanosphere. Dashed lines show polariton dispersion under assumption of "rigid" exciton with unchanged radius. The energies are all relative to the bandgap energy.

Dispersion of WEPP for different radii is shown Fig.6. Two dashed lines show dispersion of lower and upper polaritons in the "rigid" approximation i.e. disregarding the effect of VSC on the exciton radius and location. Once that coupling is introduced the lower polariton dispersion curve shifts downward. That shift, $\Delta\hbar\Omega$ is the largest (nearly 50meV) for the largest nanosphere radius $R_0 = 10nm$ because for that radius the entire exciton shell of thickness $d$ is located in the high field region, while for the smaller radii the electric field of plasmon mode drops down inside the exciton shell causing the decrease of the overlap factor $F_\kappa$ (see Fig.3). The lowering of the polariton energy makes the effective binding energy as large as $5k_BT$ where T is a room temperature, which means that it is expected to be very robust which can be beneficial for observing such phenomena as Bose Einstein Condensation at room temperature. Furthermore, the effective binding energy is larger than the linewidth of the LSP mode (~80meV) which shall assure that the lower polariton is observable. The upper polariton is not shown in Fig.6 because it is pushed up within the absorption band of CdS and thus cannot be observed.

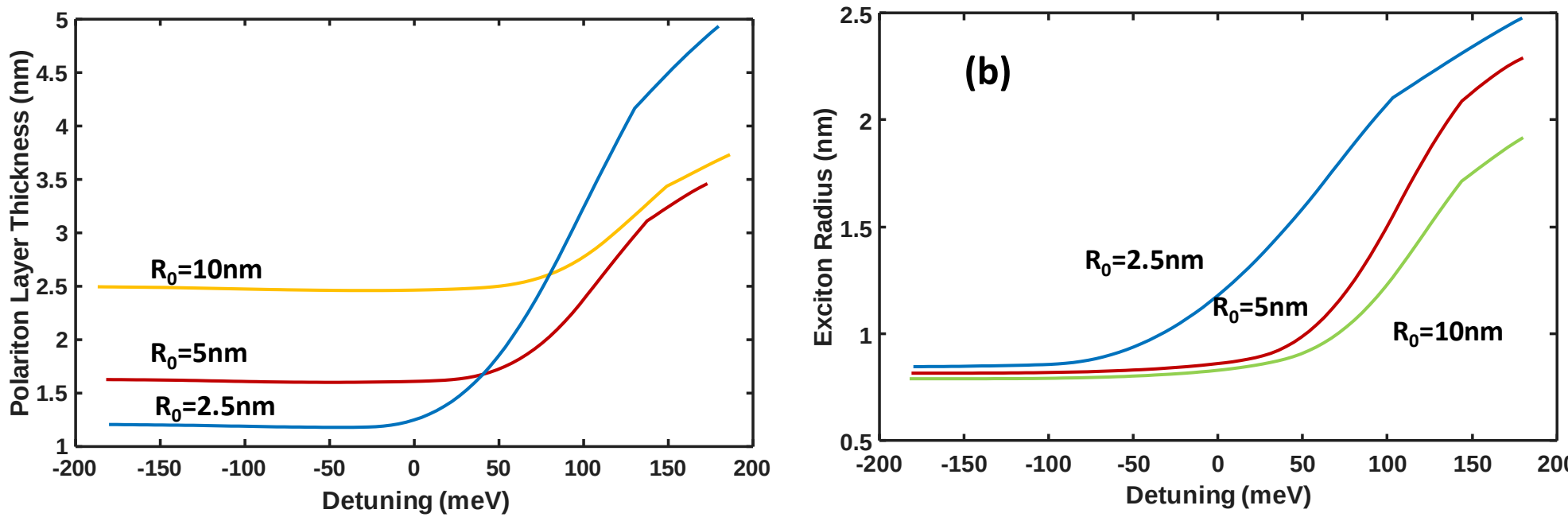

**Figure 7. (a)** Lower polariton shell thickness *d* and **(b)** excitonic radius *a* versus the energy of the "undressed" surface plasmon relative to the bandgap energy for three different radii of the Ag nanosphere. The free exciton radius is equal to 2.8nm

In Fig.7a the thickness of polariton shell $d$ is plotted versus detuning $\Delta_0$ - for lower plasmon energies the polariton gets strongly bound around the nanosphere as electron-hole pairs get attracted into the high field region where their potential energy gets lowered by emitting and re-absorbing virtual plasmons (the value of $F_\kappa$ in (16) increases according to Fig.3a). This phenomenon is akin to lowering the electron energy via emission and reabsorption of phonons when the polarons are formed [49]. It is also conceptually similar to the Lamb shift in atomic physics [50] where the energy of atomic transition in hydrogen gets reduced due to absorption and reemission of photons. At higher plasmon energies as the weight of plasmon in the lower polariton decreases and so is the product $\alpha\beta$ in (16) the lowering of potential energy of the electron-hole pairs becomes less pronounced and the electron-hole pairs spread out away from the nanoparticle surface to lower their COM kinetic energy (the value of $F_{kin}$ in (16) increases according to Fig.3b). The onset of this sharp increase in the shell thickness $d$ occurs earlier for small radii of nanoparticles where the COM kinetic energy is larger.

As one can see from Fig.7a the ratio $d / R_0$ at which maximum Rabi splitting is achieved decreases from roughly 0.56 for 2.5nm sphere to 0.27 for 10nm sphere which causes commensurate increase of the overlap factor $F_\kappa$ in Fig.3a from 0.33 to 0.48 . Therefore, Rabi splitting increases with radius $R_0$ as shown in Fig.6. When the radius $R_0$ is further increases this behavior is expected to saturate once the ratio $d/R_0$ is further reduced and reaches optimum value of 0.22. In practice though, one should be careful about further increasing $R_0$ as the polariton dimensions may exceed the coherence length, and that will reduce the oscillator strength.

Finally in Fig. 7b the dependence of exciton radius $a$ on the detuning is shown. For negative detunings the composition of polariton includes relatively small excitonic fraction $\beta^2$ (as shown in Fig.5) hence the impact of kinetic energy of relative electron-hole motion is not significant, while the small exciton radius favors the energy lowering due to virtual plasmon emission and reabsorption (Rabi energy, the last term in (16)). The exciton radius $a$ gets reduced by almost a factor of 3 relative to the "rigid" exciton radius $a_0 = 2.8nm$. The behavior is similar to the one in [13] but is more prominent due to stronger ($a^{-3/2}$ rather than $a^{-1}$) dependence of the Rabi frequency on excitonic radius. As the polariton composition changes first to more equal and then to predominantly excitonic, the kinetic energy of relative electron hole motion starts playing a more important role, while the last term in (16) decreases with the product $\alpha\beta$. Naturally, the uncoupled exciton radius then gradually increases towards free exciton radius $a_0$.

Note that decrease of the excitonic radius is the most important, and in fact the only sign of VSC as it can be easily detected by the reduction of the diamagnetic shift as it was done in [17]. Simple change of the resonant frequency between "dressed" and "undressed" exciton would be difficult to detect due to variation in samples.

## 6. Classical picture of WEPP formation

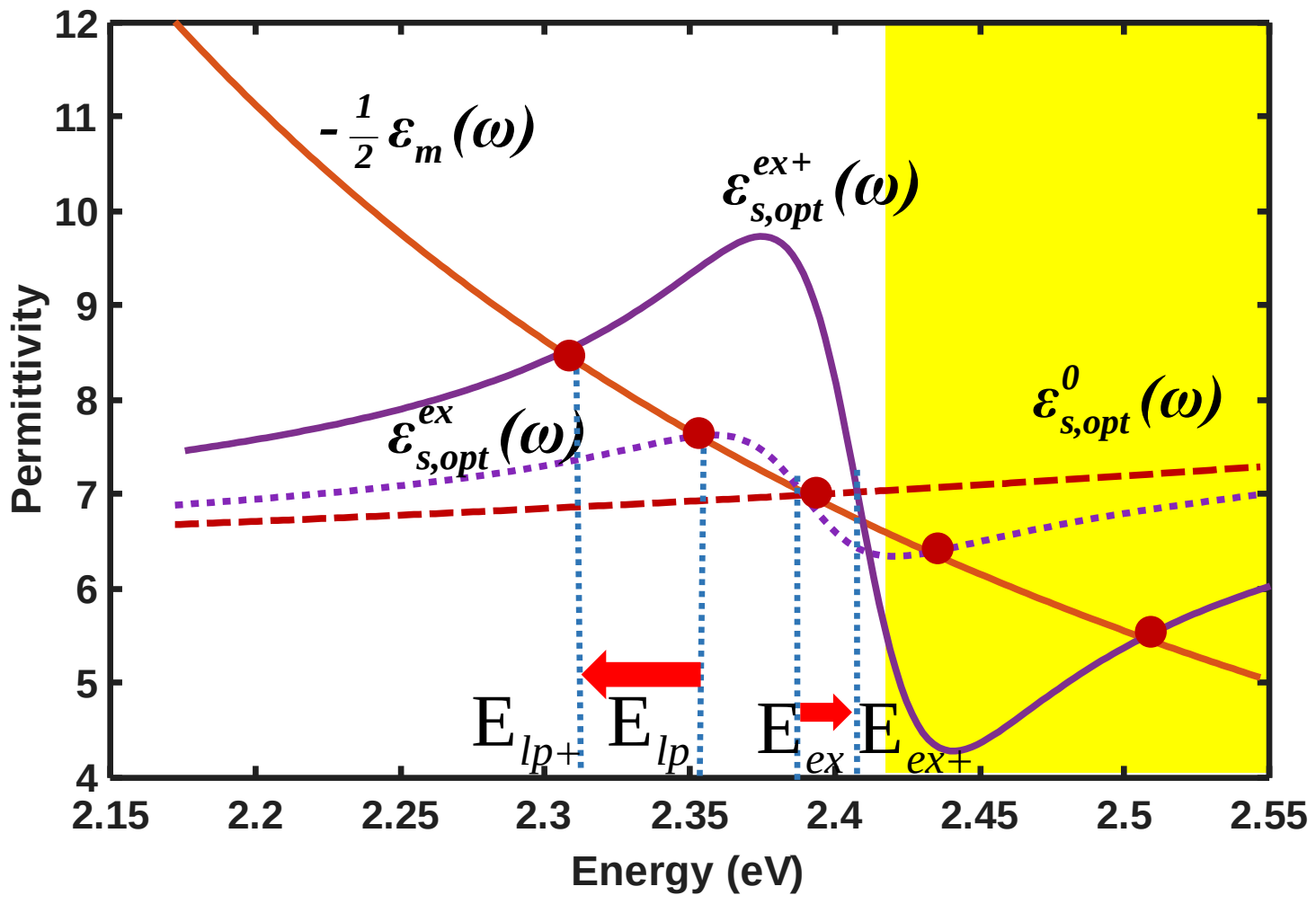

**Figure 8.** Classical interpretation of Wannier Exciton Plasmon Polariton (WEPP). Shown are the graphic solutions of Eq.(18) for the case of (A) "undressed LSP" in a semiconductor without exciton (dashed line) (B)"rigid polariton" in a semiconductor with free exciton of radius $a_0$ (dotted line) and (C) "pliable" WEPP in semiconductor with bound exciton of radius $a < a_0$ (solid line)

Before finalizing the discussion, it may be worthwhile to offer an alternative, fully classical explanation for the reduction of excitonic radius in the presence of VSC with a LSP mode. For that all one needs to do is to recall the equation defining the localized surface plasmon frequency, $\varepsilon_m(\omega_{pl}) + 2\varepsilon_{s,opt}(\omega_{pl}) = 0$, or

$$-\frac{1}{2}\varepsilon_m(\omega_{pl}) = \varepsilon_{s,opt}(\omega_{pl}) \tag{18}$$

and consider its graphic solution in Fig.8.

The left hand side of (18) (metal dispersion $-\frac{1}{2}\varepsilon_m(\omega)$) is plotted as the solid curve while the dashed curve represents $\varepsilon^0_{s,opt}(\omega)$-dispersion of the semiconductor permittivity disregarding the exciton. The intersection at point A defines $\hbar\omega_{pl} \sim 2.385eV$ - the "undressed" LSP frequency disregarding the exciton effects. Once the "rigid" exciton with radius $a_0$ is included in the dispersion (dotted curve) $\varepsilon^{ex}_{s,opt}(\omega)$ becomes anomalous around excitonic energy $\mathrm{E}_{ex} \sim 2.38$. As a consequence, the metal dispersion curve $-\frac{1}{2}\varepsilon_m(\omega)$ intersects $\varepsilon^{ex}_{s,opt}(\omega)$ at three different points of which only the first one, B, and the third one B' are stable solutions corresponding to lower and upper polaritons. Since the energy of the upper polariton B' is above the absorption edge of semiconductor, this polariton is not observable while the lower polariton can be observed at energy $\mathrm{E}_{lp} \sim 2.35eV$. Let us see what is going to happen if the excitonic radius is reduced to $a < a_0$. First of all the net difference between kinetic energy of relative electron hole motion $\mathrm{E}_{eh}$ and Coulomb attraction energy $\mathrm{E}_C$ will increase causing excitonic energy move upward to $\mathrm{E}_{ex+} \sim 2.41eV$ as indicated by the right pointing arrow in Fig.8. At the same time, reduced excitonic radius will increase the oscillator strength of exciton so that its dispersion $\varepsilon^{ex+}_{s,opt}(\omega)$ will develop more prominent anomalous Lorentzian feature around $\mathrm{E}_{ex+} \sim 2.41eV$ as shown by the solid curve which now intersects the metal dispersion at point C leading to the downward shift of the lower WEPP energy to $\mathrm{E}_{lp+} \sim 2.32eV$ as indicated the left-pointing arrow in Fig.8. The actual radius $a$ will correspond to the minimum value of $\mathrm{E}_{lp+}$ where the positive kinetic energy is balanced by negative Coulomb energy and negative coupling with plasmon energies. Thus the whole process of WEPP formation can be understood from a perfectly classical point of view, although using quantum picture greatly facilitates determination of exciton radius as it was done here.

## 7. Conclusions and Perspective

In this work we have considered very strong coupling between the Wannier exciton in semiconductor and localized surface plasmon (LSP). Using simple variational technique we have shown that once the coupling (Rabi) energy approaches the value of exciton binding energy two previously not considered phenomena ensue. First of all, the Wannier exciton gets localized in the region of high field of LSP so that a quasi-particle – Wannier Exciton Plasmon Polariton (WEPP) is formed. In addition to that the radius of the excitonic component of WEPP gets reduced by a factor of a few in comparison to free excitonic radius and the effective Rabi splitting increases accordingly. Notably, the increase in Rabi splitting beyond large broadening associated with LSP makes WEPP observable and robust even at room temperature and thus one can guardedly anticipate that Bose-Einstein condensation and polaritonic lasing may be within reach in WEPPs.

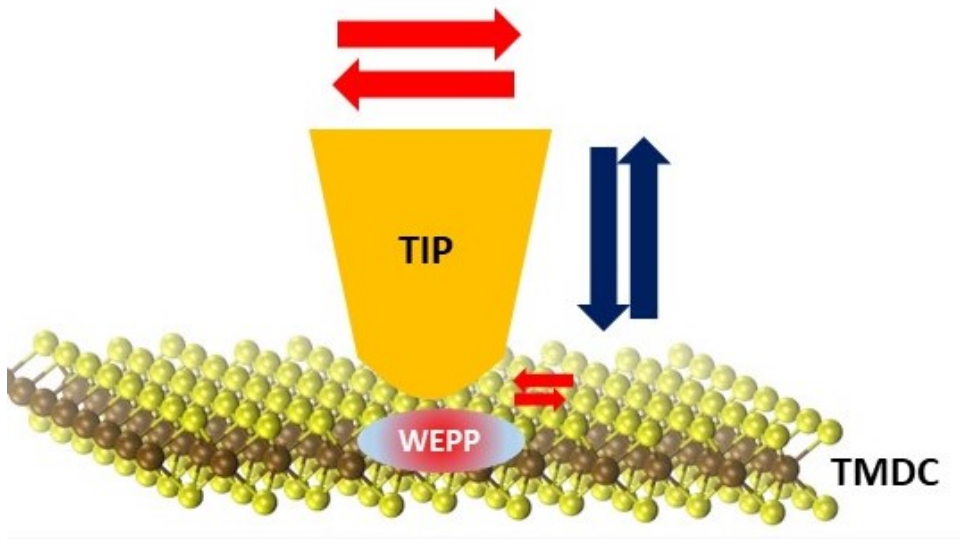

**Figure 9.** Mobile and pliable Wannier Exciton Plasmon Polariton (WEPP) in two dimensional materials (such as transition metal dichalcogenide.)

Although we have considered only the simplest and easily (almost analytically) solvable problem of WEPP bound to a spherical metal nanoparticle embedded into the semiconductor, one can consider far more interesting configurations that would allow real time manipulation of WEPPs. For instance, one can contemplate a WEPP formed by a two-dimensional exciton in a transition metal dichalcogenide material, such as $MoS_2$ or $WSe_2$ and a strong fringe (in plane) field of a metal nano-tip as shown in Fig.9. Then vertical motion of the tip will cause the change of the excitonic radius and energy in WEPP while lateral tip motion will cause the WEPP to follow the tip, just as it happens in optical tweezers, except the force moving the WEPP does not rely on optical pump and is entirely due to the vacuum field strongly enhanced in the vicinity on the tip. Furthermore, changing the geometry of the field confinement by considering, for example, field of the plasmonic dimers with their large in-plane component of the electric field, would change the shape of WEPP in a prescribed way, "sculpting" the pliable polaritonic matter.

At this point it would be prudent to avoid overhyping this new phenomenon and not to dazzle the reader with an expansive array of potential transforming applications of WEPP in every walk of life, as is regrettably often done with a great detriment to science. And yet, in my view, the remarkable physics of WEPPs, specifically their unique property of "pliability" in the engineered electro-magnetic environment, should be further explored in different geometries and material systems so that eventually applications, perhaps unanticipated, will materialize.

## Acknowledgements

The author acknowledges support from NSF grant DMR-150774 as well as endless discussions with his eminent colleague Prof. P. Noir of Johns Hopkins University that have allowed this work to crystallize in its present form.